# Flux pinning by regular arrays of ferromagnetic dots

Margriet J. Van Bael[a], Lieve Van Look[a], Kristiaan Temst[a], Martin Lange[a], Joost Bekaert[a], Ulrich May[b], Gernot Güntherodt[b], Victor V. Moshchalkov[a], and Yvan Bruynseraede[a]

[a]*Laboratorium voor Vaste-Stoffysica en Magnetisme, K.U. Leuven, Celestijnenlaan 200 D, B-3001 Leuven, Belgium.*
[b]*II. Physikalisches Institut, RWTH Aachen, D-52056 Aachen, Germany.*

**Abstract**

The pinning of flux lines by two different types of regular arrays of submicron magnetic dots is studied in superconducting Pb films; rectangular Co dots with in-plane magnetization are used as pinning centers to investigate the influence of the magnetic stray field of the dots on the pinning phenomena, whereas multilayered Co/Pt dots with out-of-plane magnetization are used to study the magnetic interaction between the flux lines and the magnetic moment of the dots. For both types of pinning arrays, matching anomalies are observed in the magnetization curves versus perpendicular applied field at integer and rational multiples of the first matching field, which correspond to stable flux configurations in the artificially created pinning potential. By varying the magnetic domain structure of the Co dots with in-plane magnetization, a clear influence of the stray field of the dots on the pinning efficiency is found. For the Co/Pt dots with out-of-plane magnetization, a pronounced field asymmetry is observed in the magnetization curves when the dots are magnetized in a perpendicular field prior to the measurement. This asymmetry can be attributed to the interaction of the out-of-plane magnetic moment of the Co/Pt dots with the local field of the flux lines and indicates that flux pinning is stronger when the magnetic moment of the dot and the field of the flux line have the same polarity.

## 1. Introduction

Recent advances in microfabrication and characterization techniques have allowed the fabrication of superconducting thin films with periodic artificial pinning arrays, such as antidot lattices [1-3] or lattices of submicron magnetic dots [4-7]. These artificial pinning arrays give rise to a huge enhancement of the critical current density ($j_c$) and the bulk magnetization ($M$) and result in the occurrence of pronounced commensurability effects at specific values of the perpendicular applied field due to the formation of stable vortex configurations in presence of the artificial periodic pinning potential. When using arrays of magnetic dots as pinning centers, the artificially created pinning potential can be expected to depend, besides on the geometry of the array, also on the magnetic properties of the dots, such as their stray field strength or the direction of their magnetic moment.

We report on the pinning properties of a thin type-II superconducting Pb film that is deposited on two different types of magnetic dot arrays. In a first type of sample, the Pb film is deposited on a square array of submicron rectangular magnetic Co dots with in-plane magnetization. By changing the domain structure of the dots, the strength of the stray field can be varied and its influence on the pinning efficiency can be determined. The second system consists of a Pb film that is grown on a square array of submicron multilayered Co/Pt dots. Due to the interface anisotropy, the magnetization in these dots is perpendicular to the substrate plane. This allows to study the direct interaction between the penetrating flux lines (FLs) in a perpendicular applied field and the magnetic moment of the dots.

## 2. Experimental

The dot arrays are prepared on $SiO_2$ substrates covered with a resist layer in which a lattice of holes is predefined by electron beam lithography. After the material deposition, the resist is removed in a lift-off procedure, leaving a lattice of dots on the substrate. A square array of rectangular Au(75Å)/Co(200Å)/Au(75Å) dots is grown in a molecular beam epitaxy (MBE) apparatus at room temperature and at a working pressure of $2 \times 10^{-10}$ Torr, using electron-beam evaporation for the Co at a rate of 0.25Å/s, and thermal evaporation from a Knudsen cell at a growth rate of 0.45Å/s for the Au layers. The thin top and bottom Au layers are added to obtain clean edges after the lift-off. These evaporation conditions result in a polycrystalline structure of the dots, as confirmed by x-ray diffraction. The lateral dimensions of the dots are $L_s \times L_l = 0.36\mu m \times 0.54\mu m$ and the period of the square lattice equals $d = 1.5$ µm. An atomic force microscopy (AFM) image of a 5µm × 5µm area of the sample is shown in Fig. 1.

A square lattice of Co/Pt multilayered dots is prepared on a $SiO_2$ substrate, making use of the same lithography and lift-off techniques. Perpendicular magnetic anisotropy can be obtained in Co/Pt multilayers because of the strong interface anisotropy, provided that the multilayer is grown in the Pt(111) direction and that the Co layer thickness is less than 12 Å [8, 9]. In order to impose the Pt(111) growth direction to the whole multilayer, a 64Å Pt buffer layer is grown, on which the [Co(5Å)/Pt(16Å)]$_{10}$ multilayer is deposited. For the deposition of the Co and Pt layers, electron beam evaporation is used in an MBE with typical deposition rates of 0.1 Å/s for both materials. An

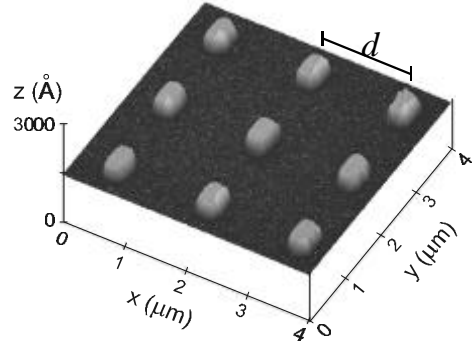

Fig. 1: Atomic force microscopy image of a 4µm×4µm area of a square lattice of rectangular Au(75Å)/Co(200Å)/Au(75Å) dots.

AFM image of the Co/Pt dot lattice is shown in Fig. 2. The lattice period equals 0.6 µm and the dots have a disk-like shape with a diameter of 0.26 µm. The local domain structure of the dots was studied at room temperature and in zero magnetic field by magnetic force microscopy (MFM) experiments on a Digital Instruments Nanoscope III system, using tapping mode and phase detection for the magnetic signal. A two-step lift mode was used to separate the topographic and the magnetic information. After characterization, the above described magnetic dot arrays are covered with a 500Å superconducting Pb film by means of electron beam evaporation at a working pressure of $10^{-8}$ Torr. By using deposition rates as high as 9Å/s, keeping the substrate cooled at 77K, a smooth continuous Pb film is obtained, completely covering the 380Å (Au/Co/Au dots) or 270Å (Co/Pt dots) high magnetic dot array. An amorphous Ge (200Å) layer is deposited on top of the Pb layer for protection against oxidation. For the Co/Pt dot array, a thin (50Å) Ge layer was also deposited underneath the Pb layer in order to prevent any proximity effect between the metallic dots and the Pb film, whereas for the other sample, the Pb film was deposited directly on top of the rectangular Au/Co/Au dots.

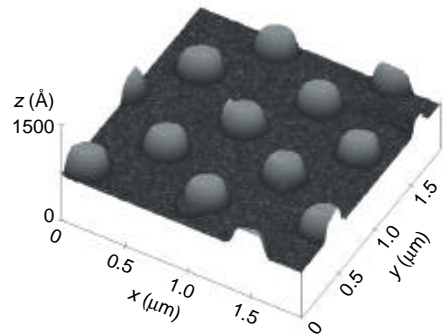

Fig. 2: AFM topograph of a 2µm ×2µm area of a square lattice of Pt(64Å)/[Co(5Å)/Pt(16Å)]$_{10}$ dots.

The superconducting flux pinning properties of these superconductor/ferromagnet hybrid structures are studied by means of SQUID magnetization measurements with the magnetic field applied perpendicular to the film plane.

## 3. Magnetic dots with in-plane magnetization as artificial pinning centers

Due to the polycrystalline structure of the Au/Co/Au dots, the magnetocrystalline anisotropy is averaged out and the magnetic properties are dominated by the shape anisotropy, resulting in a preferential magnetization direction parallel to $L_l$. MFM experiments are performed for the as-grown dot array (Fig. 3(a)), and in the remanent state after magnetic saturation along the easy axis $L_l$ (Fig. 3(b)).

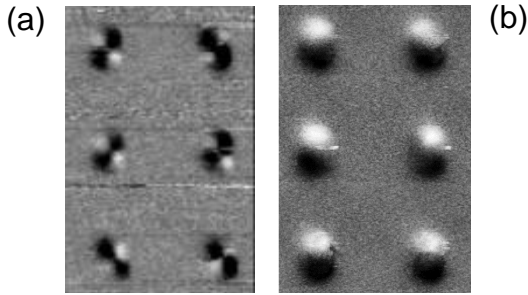

**Fig. 3:** MFM images of a 2.7μm × 4.0μm area of a square array of Au/Co/Au dots **(a)** in the as-grown state; and **(b)** in the remanent state after magnetic saturation along the long axis ($L_l$) of the dots.

In the as-grown state, the dots produce a magnetic contrast resembling a 2×2 checkerboard pattern (Fig. 3(a)), indicating a multi-domain structure with possibly two antiparallel domains. However, in the remanent state after magnetic saturation along $L_l$ in a 1 T field (i.e. well above the saturation field), the magnetic contrast of all dots consists of a bright and a dark spot, typical for a single-domain particle with in-plane magnetization [10, 11]. The existence of these two stable domain states offers the possibility to enhance the magnetic stray field of the dots by changing their domain state from multi-domain to single-domain, while keeping all other properties of the pinning array the same. This makes the array of rectangular Au/Co/Au dots an ideally suited system for studying the influence of the stray field of the dots on the pinning efficiency. For these magnetic dots with *in-plane* magnetic moment, no direct magnetic interaction between the local field of the FLs in a perpendicular applied field and the magnetic moment of the dots is expected. After deposition of the superconducting Pb layer, we performed magnetization measurements of the hybrid superconducting/magnetic system before and after the dots are magnetized. The critical temperature of the sample is $T_c = 7.16K$. Fig. 4 shows the upper branch of the magnetization loop at $T/T_c = 0.97$ versus $H/H_1$ where $H_1$ corresponds to the first matching field, defined as the field value at which the density of FLs equals the density of dots:

$$\mu_0 H_1 = \phi_0/(1.5\mu m)^2 = 0.92 \text{ mT},$$

with $\phi_0$ the superconducting flux quantum. The presence of the dot array results in a huge enhancement of the magnetization and the critical current density ($M(H) \sim j_c(H)$) when compared with a reference Pb(500Å) film without dots (full line in Fig. 4). For temperatures close to $T_c$, pronounced anomalies are observed at certain integer and rational multiples of $H_1$, e.g. a sudden drop of $M(H)$ at $H/H_1 = 1$, and a maximum at $H/H_1 = 2$ and 3 in Fig. 4. These matching anomalies are caused by the commensurability of the imposed square pinning potential (the array of dots) and the lattice of FLs at specific fields, which results in abrupt changes of the stability of the flux line lattice (FLL) as the applied field is changed. The strong enhancement of $j_c$ and the matching effects are observed for the magnetized as well as for the non-magnetized dots (see Fig. 4) and indicate that the dots create a strong periodic pinning potential for the FLs in the superconducting Pb film. The pinning potential is created due to a combination of several contributions, including (i) the geometrical modulation of the Pb film because of the underlying dots; (ii) there is a possible influence of the proximity effect between the metallic dots and the superconducting Pb layer, locally lowering the superconducting order parameter; and (iii) the influence of the local stray field of the dots, which locally weakens or destroys superconductivity. By comparing the magnetization results obtained before (Fig. 4, filled symbols) and after (open symbols) the dots are magnetized, we can focus on the influence of the magnetic stray field. From Fig. 4, it is clear that after magnetizing the dots (i) $M$ is further increased, (ii) the matching anomalies are more pronounced, and (iii) additional matching anomalies appear (see e.g. at $H/H_1 = 3/2$). These observations indicate that *the pinning is most efficient in the broadest field range for magnetic dots with the largest stray field strength.* Similar results were found for a triangular array of magnetic Au/Co/Au dots as artificial pinning array, from SQUID magnetization experiments [6] as well as by measuring the critical current density directly by means of electrical transport measurements in a magnetic field [7].

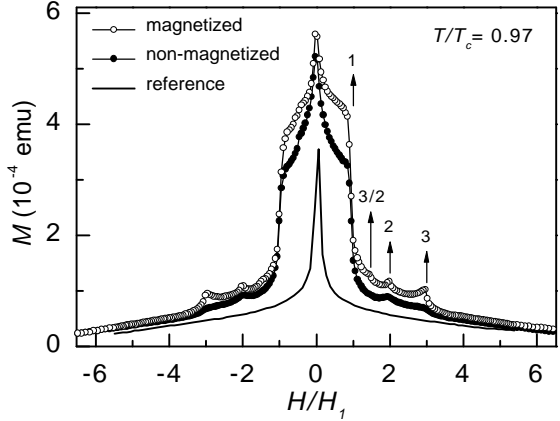

Fig. 4: Upper half of the magnetization loops $M(H/H_1)$ at $T/T_c = 0.97$ of a Pb(500Å) film on a square array of Au/Co/Au dots ($d = 1.5$ µm) before (filled symbols) and after (open symbols) the dots are magnetized, and for a reference Pb(500Å) film without dots (full line).

Magnetization curves of the Pb film on the square array of Au/Co/Au dots have been measured at different temperatures, revealing that the matching anomalies are less pronounced as the temperature is lowered, and even disappear at $T/T_c = 0.85$. Very close to $T_c$, on the other hand, matching anomalies are also observed at a significant number of rational multiples of $H_1$. This temperature dependence can be understood when considering that the long-range repulsive interaction of the FLs and a uniform flux density in large parts of the sample are crucial factors for the observation of matching anomalies. These requirements are only fulfilled at temperatures close to $T_c$ where the penetration depth $\lambda(T)$ is large compared to the distance between the FLs and where the flux gradient in the sample is small. The suggested stable vortex configurations giving rise to the observed anomalies at integer and rational multiples of $H_1$ are shown schematically in Fig. 5, where the black dots and the open circles represent the pinning centers and the FLs, respectively. In these configurations, it was assumed that only one flux quantum can be pinned at each pinning center, and no multi-quanta FLs are formed at the dots. This is confirmed by the shape of the magnetization curves for $T$ very close to $T_c$, specifically by the sudden drop of $M$ at $H_1$ (see Fig. 4), which reveals that at $|H| > H_1$ *interstitial* FLs are present in the sample. These interstitials are not pinned at the dots and are therefore characterized by a higher mobility, resulting in the sudden decrease of $M$ and $j_c$ at $H_1$. At lower temperatures, however, this is not valid and multi-quanta flux lines can be pinned at the pinning centers, as was shown in scanning Hall probe microscopy experiments on a similar sample, in which no interstitials are observed [12]. Most of the specific stable configurations of the FLL in a square periodic pinning potential that are shown in Fig. 5 have also been found by molecular dynamics simulations [13], or have been directly visualized by local observation techniques such as Lorentz microscopy [14].

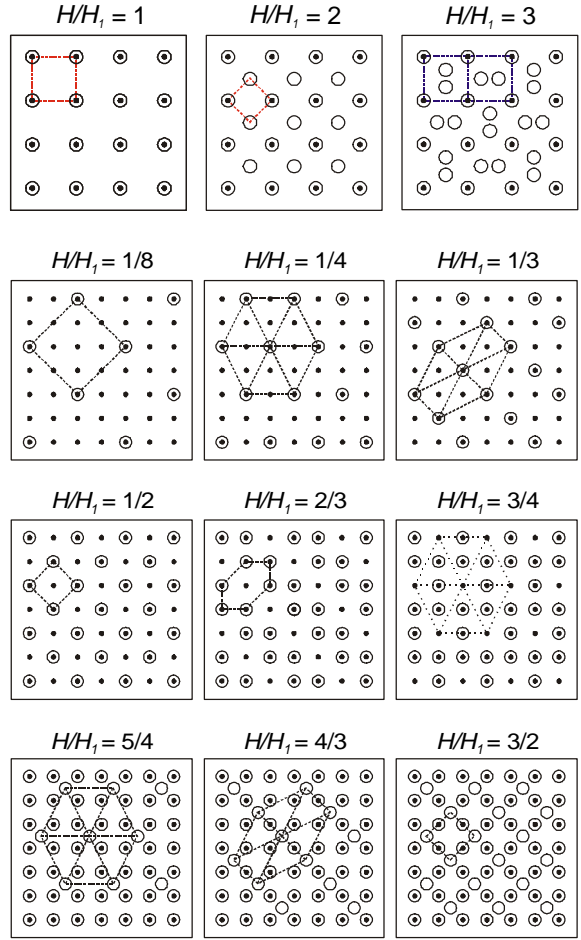

Fig. 5: Schematic presentation of the suggested stable flux line lattices at integer and rational multiples of the first matching field $H_1$. The black dots and the open circles indicate the pinning centers and the flux lines, respectively. The dashed lines are guides to the eye, showing the symmetry of the flux line lattice.

## 4. Magnetic dots with out-of-plane magnetization as artificial pinning centers

A square array of Co/Pt dots with *out-of-plane* magnetization is used as a magnetic pinning array for a superconducting Pb film that is deposited on top. In contrast to the Au/Co/Au dots, which have their magnetic moment in the substrate plane, the FLs in a perpendicular applied field can interact with the perpendicular magnetic moment of the Co/Pt dots. In this experiment, our aim is to determine the influence of the mutual orientation of the magnetic moment of the dots (*m*) and the FLs on the pinning efficiency, i.e., can we observe a clear difference in pinning strength between the case when *m* is parallel with the applied field and the penetrating FLs (Fig.

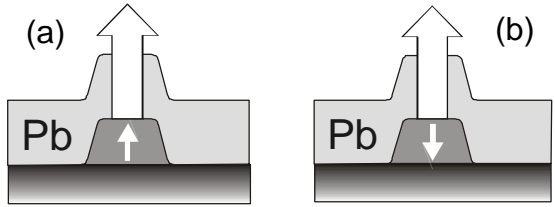

Fig. 6: Schematic drawing of the cross-section of a Co/Pt dot covered with a Pb film, presenting a flux line (large arrow) in a perpendicular field, pinned at a magnetic dot with a out-of-plane magnetic moment (small arrow), showing the parallel (a) and anti-parallel (b) alignment of the field of the flux line and the magnetic moment of the dot.

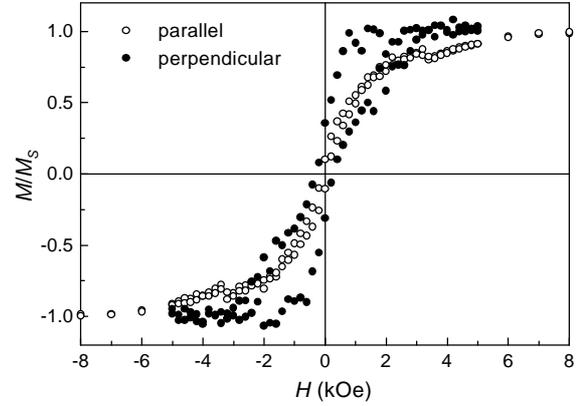

Fig. 7 SQUID hysteresis loops of a square array of Pt(64Å)/[Co(5Å)/Pt(16Å)]$_{10}$ dots at $T = 5$K with the applied field parallel (open symbols) and perpendicular (filled symbols) to the substrate plane.

6(a)) and the case where *m* is anti-parallel with the FLs (Fig. 6(b)).

The magnetic properties of the square lattice of Pt(64Å)/[Co(5Å)/Pt(16Å)]$_{10}$ dots are determined before deposition of the superconducting Pb layer. SQUID magnetization measurements (Fig. 7) of the dot lattice with the field parallel and perpendicular to the sample plane confirm that the easy axis of magnetization is indeed perpendicular to the substrate plane, as can be seen from the lower saturation field in the perpendicular configuration. Note also that a net positive or negative magnetization value, corresponding to about 35% of the saturation magnetization, is found in the remanent state ($H = 0$) of the dot array after saturation in a positive (up) or negative (down) perpendicular field, respectively. This indicates that after saturation in a positive (negative) perpendicular field, most of the dots will have a positive (negative) out-of-plane magnetic moment, i.e. pointing upwards (downwards). MFM measurements at room temperature and in zero field have been performed on the as-grown Co/Pt dot array to get an idea about the domain structure of the dots. The resulting MFM image displayed in Fig. 8 was recorded during the first scan over that particular area and was scanned downwards from the top of the figure. Most of the dots produce a uniform dark or bright MFM signal, indicating a single-domain structure with out-of-plane magnetization. However, the fact that the majority of the dots give a dark image suggests that the stray field of the MFM tip switches the magnetic moment of the dots while scanning over them. Traces of an initially bright MFM signal, suddenly changing to dark, can be found in many of the dots in Fig. 8. We can therefore only conclude that at least part of the dots have a single domain magnetic structure, with a random distribution of dots having a positive or negative magnetic moment in the as-grown array. From the magnetic characterization of the Co/Pt dots, we may assume that the Co/Pt dots indeed have an *out-of-plane* magnetic moment *m*, which can be oriented in the positive or negative direction after magnetic saturation. After deposition of the superconducting Pb film, SQUID magnetization measurements are performed below the critical temperature ($T_c$ = 7.20 K) on the as-grown dot array, and after saturation of the dots in a positive and a negative perpendicular field. Figure 9 shows the resulting hysteresis loops at $T/T_c$ = 0.97 with the dots magnetized in a positive (open symbols) and a negative (filled symbols) perpendicular field. The first matching field for this square array with period 0.6μm equals $\mu_0 H_1 = \phi_0/(0.6\mu m)^2$ = 5.74 mT. A pronounced asymmetry of the *H*>0 and *H*<0 parts can be observed in both *M(H)* loops. Clear matching effects are only observed in that part of the curve where the field polarity (i.e. the field direction of the FLs) is aligned parallel with *m* of the dots, namely in the *H*>0 part for *m*>0 and in the *H*<0 part for *m*<0. In these parts, the magnitude of $M(H) \sim j_c$ is also markedly larger than in the part where the applied field and *m* are anti-parallel. This indicates that a lattice of dots with *m*>0 provides a much stronger periodic pinning potential for positive FLs (*H*>0) compared to dots with *m*<0, and vice versa. For the flux line configurations giving rise to the observed matching effects, we can again refer to Fig. 5. The inset of Fig. 9 shows the *M(H)* result for the as-grown dot array, i.e. in which dots with *m*>0 and *m*<0 are randomly distributed over the square array. No matching effects could be found in this case, neither for lower temperatures nor for *T* even closer to $T_c$. In order to explain these results, the direct magnetic interaction of the dots and the FLs should be taken into account. Many different contributing interaction terms play a role in the total magnetic interaction, involving the magnetic moment of the dots (*m*), the local magnetic field of the FLs (*h*), the screening currents of the FLs, screening currents around the dots, the stray field of the dots, etc. A simplified picture, taking into account only the interaction energy *E* of the magnetic moment of the dots and the local field of the FLs [5]:

$$E = -\mathbf{m}\cdot\mathbf{h}, \tag{1}$$

can however qualitatively explain our observations, and can therefore also be considered as the most important of these magnetic interaction terms. According to Eq. 1, pinning of a flux line at a magnetic dot with parallel *m* results in an energy reduction, whereas an energy enhancement is obtained for the pinning of a flux line at a dot with anti-parallel **m**. This results in a stronger attraction (i.e. better pinning) of FLs to dots with **m** parallel to **h**, and a repulsive interaction contribution for the case where **h** and **m** are anti-parallel. The total pinning potential is the sum of the above discussed magnetic interaction (Eq. 1) and all other possible contributions, like e.g. the geometric modulation of the Pb film following the topography of the 274 Å high dots, the effect of the stray field of the dots, etc. This total pinning potential will have deep minima at the positions of the dots with **m** parallel to the applied field (and hence to **h**) and will have only very shallow energy minima or even maxima at the position of the dots with **m** anti-parallel to *H* and **h**. After saturation of the dot array in a positive (or negative) perpendicular field, a strong periodic pinning potential is *only* present when *H*>0 (or *H*<0, respectively). In the Pb film on the as-grown dot array, the pinning potential landscape will not be periodic due to the random distribution of dots with **m**>0 and **m**<0. The lack of a periodic pinning potential explains the fact that no matching effects are observed in the Pb film on the as-grown dot array.

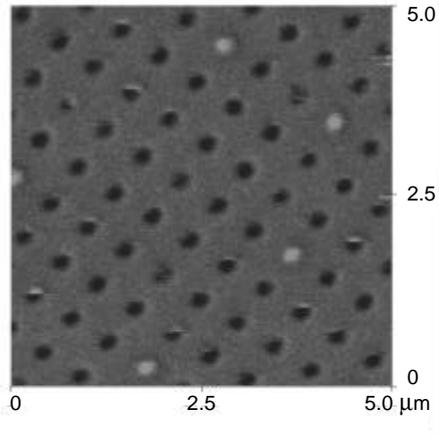

Fig. 8: MFM image of the as-grown square lattice of Pt(64Å)/[Co(5Å)/Pt(16Å)]$_{10}$ dots, scanned downwards from the top of the picture.

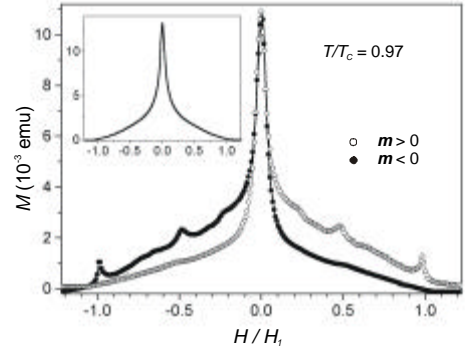

Fig. 9: Upper part of the magnetization curves versus $H/H_1$ of a Pb film on a square array of Co/Pt dots (with $H_1$ = 5.74 mT the first matching field), measured at $T/T_c$ = 0.97 after magnetizing the dot array in a positive (open symbols) and a negative (filled symbols) perpendicular field. The inset shows the $M(H/H_1)$ result for a Pb film on the as-grown Co/Pt dot array.

## 5. Conclusion

In conclusion, we have studied the pinning properties of a superconducting Pb film on two different types of *magnetic* artificial pinning arrays. Both types of dot arrays act as strong artificial pinning lattices of which the pinning potential depends strongly on the magnetic properties of the dots. A significant enhancement of the critical current density and pronounced matching effects are observed at several integer and rational multiples of the first matching field, which are attributed to the formation of stable vortex configurations in the imposed artificial square pinning potential. By controlling the domain structure of the Au/Co/Au dots, the influence of the stray field strength on the pinning potential was determined, showing that increasing the stray field of the magnetic dots enhances the pinning efficiency. When using a lattice of Co/Pt dots as pinning array, the out-of-plane magnetic moment of the dots interacts directly with the FLs in a perpendicular applied field. This results in a pronounced asymmetric *M*(*H*) behavior for parallel and anti-parallel alignment of the magnetic moment of the dots and the local field of the FLs and allows to conclude that flux pinning is much stronger when the magnetic moment of the dots and the local field of the FLs have the same polarity.

## 6. Acknowledgement

This work was supported by the Fund for Scientific Research – Flanders (FWO), the European ESF VORTEX Program, and by the Belgian Inter-University Attraction Poles (IUAP) and the Flemish Concerted Research Actions (GOA) Programs. MJVB and KT are Postdoctoral Research Fellows of the FWO.